\documentclass[nohyper,12pt,letterpaper]{JHEP}
\usepackage{epsfig}

\newfont{\frak}{eufm10 scaled 1200}

\newfont{\Bbb}{msbm10 scaled 1200}     
\newcommand{\mathbb}[1]{\mbox{\Bbb #1}}
\DeclareSymbolFont{AMSa}{U}{msa}{m}{n}
\DeclareSymbolFont{AMSb}{U}{msb}{m}{n}
\let\Box\relax
\DeclareMathSymbol{\Box}{\mathord}{AMSa}{"03}

                                                \def\beq{\begin{eqnarray}}
\def\eeq{\end{eqnarray}}
\def\bea{\begin{eqnarray*}}
\def\eea{\end{eqnarray*}}

\title{Dark Energy in Perturbative String Cosmology}

\author{T. Banks\thanks{On leave from Rutgers University.}, M. Dine\\
  Department of Physics and Institute for Particle Physics\\
  University of California, Santa Cruz, CA 95064\\
E-mail: \email{banks@scipp.ucsc.edu, dine@scipp.ucsc.edu}}

\abstract{The apparent observation of dark energy poses problems for string
theory.  In de Sitter space, or in quintessence models, one cannot
define a gauge-invariant S-matrix.  We argue that eternal
quintessence does not arise in weakly coupled string theory, but
point out that it is difficult to define an $S$-matrix even in the
presence of perturbative potentials for the moduli. The solutions
of the Fischler-Susskind equations all have Big Bang or Big Crunch
Singularities.  We believe that an S-matrix (or S-vector) exists
in this context but cannot be calculated by purely perturbative
methods. We study the possibility of metastable de Sitter vacua
in such weakly coupled scenarios, and conclude that the S-matrix
of the extreme weak coupling region cannot probe de Sitter
physics.
 We also consider proposed explanations of the dark
energy from the perspective of string theory, and find that most
are implausible.  We note that it is possible that the axion
constitutes both the dark matter and the dark energy.}


\received{???????? ?st, 2000} \accepted{???????? ?th, 1998}

\preprint{\hepth{0106276}\\SCIPP-2000/04}
\begin{document}


\section{Introduction}

Accelerating universes are not compatible with the conventional
setup of string theory.   The appearance of a cosmological horizon
in many models incorporating rolling scalar fields signals the
absence of a completely gauge invariant
S-matrix\cite{tbf}\cite{fetal}\cite{hks}. This poses a
phenomenological challenge for string theory, since our universe
appears to be accelerating \cite{acc}.   One may ask however if
it can lead us to an inconsistency of the string theoretic
formalism itself. That is, supersymmetric vacua of string theory
have exactly massless moduli fields at tree level.  There are a
variety of situations in which we believe that we can break SUSY,
either at tree level, or through low energy nonperturbative
effects, in a controllable manner. That is, SUSY breaking
generates a potential for the moduli, which attracts the system
to the weak coupling regime, where the cosmological constant
vanishes.  One can then imagine setting up a scattering theory in
which we choose a solution of the effective equations of motion
in which the dilaton is in the extreme weak coupling regime both
in the infinite past and the infinite future. The asymptotic
states are then those of freely moving string excitations, and we
can imagine computing the S-matrix for scattering from some free
particle state in the past to a different one in the future.
Fischler and Susskind \cite{fs} have provided a prescription that
purports to obtain well defined vertex operator correlation
functions in the case that the potential is generated in
perturbation theory.   These are presumably the S-matrix elements
in question.  We will use the phrases Fischler-Susskind Cosmology
and Cosmological S-matrix (FSC and FSCS-matrix) to describe both
perturbative and nonperturbative scenarios of this kind.

The papers of \cite{fetal}\cite{hks} pose a potential problem for
this program.  They show that in many models with rolling scalar
fields there is a cosmological horizon and no sensible S-matrix
exists. In this paper we will show, following old work of
Brustein and Steinhardt \cite{bs} that this problem does not
occur in FSC.  No plausible FSC solution has a horizon. However,
we will point out that all such solutions have singularities of
the Big Bang or Big Crunch type.  Thus, it is {\it not} true that
we have a controllable weak coupling string calculation in this
scenario.

Our analysis is done at the level of low energy effective field
theory and it is barely possible that $\alpha^{\prime}$
corrections eliminate the singularity.  We argue that this is
unlikely to be true, by using the holographic principle.

Nonetheless, we believe that in all of these situations, an
S-matrix or S-vector\cite{tbf}\cite{w} exists, though we may need
the full nonperturbative definition of the theory to describe the
initial state (in the case of a Big Bang).

This leads us to another question.  Suppose that, by utilizing a
variety of fluxes\cite{joeetal}, one succeeds in calculating a
potential for moduli that stabilizes them in a regime where string
perturbation theory is apparently accurate. Suppose further that
the value of the potential at the nontrivial minimum is
nonnegative.   How does one describe the properties of the stable
or metastable state at this minimum in terms of the original
weakly coupled string theory from which the potential was derived.
That is, ignoring the problems with the Big Bang singularity, can
one set up an initial state that will probe the properties of the
nontrivial minimum?

We argue, using the results of \cite{gf} and \cite{isovac} that
this is unlikely to be the case.   These papers argued that in an
asymptotically Minkowski spacetime, generic attempts to access a
DeSitter or isolated Minkowski minimum, lead instead to the
formation of a black hole.  The same black hole can be created in
ways that have nothing to do with the nontrivial minima (the No
Hair theorem).   According to Black Hole Complementarity
\cite{thsut}, physics as seen by observers inside the black hole
is described by operators (including the time evolution operator)
that do not commute with any of the observables at infinity. Thus
there is no way for the scattering matrix to encode information
about physics at the nontrivial minimum, as it would be
experienced by an observer who thought that he/she was in
DeSitter space.

Although FSC spacetimes are not asymptotically flat, they are, in
Einstein frame, Friedmann-Robertson-Walker (FRW). Thus, the same
conclusions can be drawn in this case (modulo lingering
uncertainties about the meaning of initial conditions at the Big
Bang).  The transformation between Einstein and string frames is
asymptotically singular in the weak coupling regime (and both
metrics are singular at the Big Bang).   The proper framework for
any attempt to construct a perturbative S-matrix is the string
frame, because the string frame metric is what appears in the
vertex operator construction of scattering amplitudes.
Nonetheless, we argue that the conclusions about the FSCS-matrix
that we draw from the Einstein frame analysis are completely
valid.

Stepping away from the larger conceptual issues that are the main
subject of this paper, we examine a number of proposals for the
dark energy.  We conclude that existing proposals require
phenomena far different than any known in string theory, or
remarkable coincidences beyond the cancellation of the
cosmological constant and the near equality of the dark matter and
dark energy densities.  We discuss the proposal of
\cite{tbfolly}, arguing that it is consistent with the observed
facts.  We also note that the QCD axion could plausibly constitute
both the dark matter and the dark energy.

\section{No Cosmological Horizons in Perturbative String Theory}

Some years ago, Brustein and Steinhardt \cite{bs} argued that the
potentials expected from SUSY breaking scenarios in weakly coupled
string theory could not give rise to inflation, nor indeed to any
asymptotically accelerating expansion of the universe. The
argument was given in the context of a single scalar field, and is
easy to recapitulate. We will work in Einstein frame because that
is the frame in which horizon areas are a measure of entropy.

The equations of motion of a minimally coupled (canonically
normalized) scalar field and gravity, assuming a $d+1$
dimensional, flat FRW universe, are: \beq {H^2 \equiv
({\dot{a}\over a})^2 = {2 \kappa^2 \over d (1-d)} E \equiv
{\dot{\phi}^2 \over 2} + V(\phi )} \label{eoma} \eeq \beq
{\ddot{\phi} + d H \dot{\phi} + V^{\prime} (\phi ) = 0}
\label{eomb} \eeq where $\kappa^2 = 8 \pi G_N$ is the
gravitational coupling. Assuming an expanding universe, these can
be written as a single equation for $E$ as a function of $\phi$:
\beq {E_{\phi} = - 2 \sqrt{d \over d-1}\sqrt{2 E(E - V)}}
\label{EQN} \eeq

We are in a situation where the potential is positive, and
asymptotes to zero at $\phi = \infty$.   The question of whether
there is a cosmological horizon depends only on how fast $V$
asymptotes to zero.  To see this, note that the coordinate of the
horizon is given by

\beq
{R_H = \int {dt\over a} = \int d\phi [{1\over
a\dot{\phi}}] = \int d\phi [a \sqrt{2(E - V)}]^{-1}}
\label{horcoor}
\eeq

The scale factor, $a$, is given by:
\beq
{a = exp[\int d\phi  \sqrt {E\over d(d-1)(E - V)}]}
\label{scalefactor}
\eeq

If the energy is asymptotically dominated by the potential, then
$a$ will diverge at large $\phi$, much more rapidly than $E
\rightarrow V$.  Then the integral defining $R_H$ will be finite
for all times, and there is a cosmological horizon. Nothing beyond
the coordinate distance $R_H$ will be visible to a local observer.

Now suppose $V$ vanishes more rapidly than an exponential.  We
claim that $E \gg V$ asymptotically.  Indeed, neglecting $V$ in
\ref{EQN} we see that $E$ vanishes exponentially and the
approximation is self-consistent.  Conversely, if $V$ vanishes
less rapidly than an exponential then so must $E$.   Thus $E$
must asymptote to $V$, $E = V + \Delta$, with

\beq{\Delta_{\phi} \simeq -2 \sqrt{2d\over
d-1}\sqrt{V\Delta}}\label{corr}\eeq

Consider finally the case where $V = e^{-\alpha\phi}$ and define $
E \equiv r V$.  Then:
\beq
r_{\phi} - \alpha r = - 2\sqrt{d \over d-1} r\sqrt{(1-r^{-1})}
\label{rqn}
\eeq
with the condition $ r \geq 1$.  This equation has a fixed point,
which automatically obeys the bound on $r$.   Otherwise, its
asymptotic behavior is dominated by large $r$.   In either case,
$E = C e^{-\beta\phi}$ with
\beq
{\beta = \pm 2d \sqrt{d \over d-1}}
\label{inf}
\eeq
when $r \rightarrow\infty$ and
\beq
{\beta = \alpha},
\label{fp}
\eeq
for the fixed point solution.  It is easy to see that there will
be a horizon only for the fixed point case, and only if
${2 \over \alpha(d-1)} >1$.

The result of Brustein and Steinhardt is now easy to understand.
In string theory, the coupling is an exponential of a canonically
normalized dilaton field.   Thus, any nonperturbative potential
will be an exponential of an exponential and there will be no
horizon.   For a potential generated at one loop, we have to do a
bit more work, to figure out the exponent in terms of the
canonical dilaton in Einstein frame.   However, this is easy and
even the one loop potential violates the bound on $\alpha$
implied by the existence of a horizon.  In ten dimensions, for
example, if the potential is generated at one loop, $\alpha= {5
\over 4 \sqrt{2}}$.  In four dimensions, the situation is more
complicated, since there are typically several moduli.  In an
asymmetric orbifold compactification with no geometric moduli, or
in other compactifications with all of the moduli but $S$ frozen,
$g^2 = e^{\phi \over \sqrt{2}}$, and a one loop potential would
have $V = c g^2$. This is an exponential of the dilaton vanishing
more rapidly than the bound.

One may question whether this result is truly general.  The
tachyon free SUSY violating heterotic string in ten dimensions
provides an example with only a dilatonic modulus.  More general
FSC models will have many moduli.  There is a metric on moduli
space and the free motion of homogeneous modes on the moduli space
 is chaotic\cite{gmh}.  Furthermore, the coupling of the
 homogeneous and inhomogeneous modes produces an instability for
 production of the latter\cite{bbmss}.

 Fortunately, none of these questions affect the existence of a
 horizon, since it is a phenomenon which occurs only
 asymptotically, and only if the motion of the moduli is, in the
 end, dominated by the potential energy.   Because of the potential,
 even the chaotic system will
 eventually reach the asymptotic region (remember, according to
 the ground rules of FSC models, the potential is everywhere
 positive and has no minima except at infinity).  Furthermore, the
 instability of \cite{bbmss} shuts off for small velocity and will
 not affect the late stages of motion if there is a horizon.

 Now note that the Friedmann equation and the equation
 for $E$ as a function of time, $ \dot{E} = - d H \sqrt{2(E - V)}$
 , are still valid, with $E$
 given by $E = {1\over 2} \dot{M}^i G_{ij} (M) \dot{M}^j + V(M)$.
 Any solution of the equations of motion for the moduli, will
 follow some trajectory through moduli space.  Define the path
 length variable $Z = \int dt \sqrt{\dot{M}^i G_{ij} \dot{M}^j}$.
 Then $E = {1\over 2} \dot{Z}^2  + V(M(Z))$.  We now have a
 problem of a single variable again, with $Z$ playing the role of
 $\phi$.  So we need only know the behavior of $V$ along the
 asymptotic trajectory in moduli space, as a function of the path
 length.

 Asymptotic directions in moduli space correspond to $d+1$
 dimensional coupling going to zero (because we are talking about FSC
 models), perhaps combined with a blowup of some internal
 dimensions.  Models with blowing up internal dimensions are
 harder to analyze, because the asymptotic effective field theory
 has a larger number of dimensions than the initial
 theory.  For trajectories where blowup does
 not occur, our previous analysis holds.   We have only cursorily
 examined other models and found no examples with horizons, but
 have stopped short of an exhaustive study because we are not
 completely sure how to interpret the S-matrix when some
 dimensions become large asymptotically.

 \section{\bf Bangs and Crunches}

Saving the universe from quintessence has turned out to be a
simple task in FSC models.  Saving it from the singularity theorems
of Penrose
and
Hawking\cite{singularities} is quite another matter. Indeed, asymptotically,
all FSC models approach $p = \rho$ FRW universes, since the
kinetic energy dominates the potential.  Thus, they contain a
singularity either in their past or their future (note that the
$d$ spatial dimensions are not compactified and we cannot use
dualities to remove singularities).   This might not be so bad if
the singularity of the asymptotic solution in the future, was in
the past, while that of the asymptotic past solution was in the
future.   Then we could hope that the complicated intermediate
dynamics somehow removed the singularity.   This cannot happen.

The desired state of affairs requires expansion in the remote
future and contraction in the remote past.  But the Friedmann
equation states that $H$ can change sign only at zeroes of the
energy, and the energy can only vanish if $\phi = \infty$ and
$\dot{\phi}$ approaches zero there (remember again that the
philosophy of FSC models requires us to remain in regions where
the asymptotic form of the potential is valid).

We can try to fix this up with a positive spatial curvature.  Now
$H$ can vanish for positive energy.   It is easy to see that if
the curvature term is large enough when $H$ vanishes then
contraction truly turns into expansion.   Unfortunately, there is
a fly in the ointment.  Asymptotically, the field energy vanishes
like $a^{-2d}$ while the curvature falls only like $a^{-2}$. Thus,
we can have neither contraction from infinite size in the past nor
expansion to infinite size in the future.  The universe undergoes
a Big Crunch, and began with a Big Bang.

After contemplating these disasters for a few minutes, one
realizes that one was ``doomed from the start".  Our models all
satisfy the conditions of the Hawking-Penrose singularity
theorems.  The generality of those results leads us to conclude
that there is no escape within the realm of low energy effective
field theory.   Are there stringy loopholes?

\subsection{\bf Can We Blame the Frame?}

Since our solutions evolve to regions of weak coupling, we can
phrase the problem of solving the string equations in terms of
solving the $\beta$-function equations for a two dimensional
conformal field theory.  One potentially mitigating feature of
these equations is that they are most naturally written, not in
the Einstein frame, but in the string frame.  In the string
frame, the curvatures are smaller than in the Einstein frame by a
power of the coupling constant.  In this section, we will see,
however, that even the string frame curvature is too singular to
permit a perturbative solution of the beta function equations. We
will see that this feature holds independent of the detailed form
of the potential (e.g. whether it arises at one or two loops, or
non-perturbatively).  Thus if the problem has a solution, it
cannot be found perturbatively in the $\alpha^\prime$ expansion.
Conceivably, one can find an exact conformal field theory which
is non-singular.  It is interesting that, because the asymptotic
behavior of the curvature (and the dilaton) in the singular
region is independent of the potential, this problem is likely to
either have no solution at all, or a solution which is universal.

To determine the behavior of the potential, it is helpful to look
at the solutions of the (Einstein frame) equations in more detail.
Consider first the case of ten dimensions, and  suppose we
consider a theory such as one of the ten dimensional, non-supersymmetric,
non-tachyonic
string theories, which develops a potential at one loop.

In terms of the canonically normalized dilaton field, $D$, the
potential has the form:
\beq
V= e^{- \alpha D}
\eeq
while the coupling is given by
\beq
g^{-2} = e^{-\gamma D}
\eeq
One can determine the asymptotic behavior of the fields by means
of the procedure outlined above.  For the case of an expanding
universe, one finds:
\beq
t \rightarrow -\infty:  a \rightarrow e^{\sqrt{1 \over d (d-1)}\phi}
\eeq
so the curvature, in the Einstein frame, behaves as
\beq
{\cal R} = -d(1-d)({\dot a \over a})^2
\\ ({2 \over d(d-1)} e^{ \sqrt{d \over d-1} \phi}.
\eeq
On the other hand, the coupling behaves, in ten dimensions, as
\beq
g^2 = e^{-{\phi \over \sqrt{2}}}
\eeq

So even in the string frame, the curvature blows up.
Again, note that the problem is universal; it does not depend on the
details of the potential.  In other dimensions, the results are
similar.  For example, in four dimensions, if we freeze the volume
modulus, $g$ behaves identically, whereas $d=3$ in the formula for
the behavior of $a$.

We conclude that the Big Bang singularities of FSC solutions
cannot be removed by the conformal transformation to string frame.
It is conceivable that exact conformal field theories could be
found, which removed the singularity found at lowest order in the
$\alpha^{\prime}$ expansion.   However, the results of \cite{bfm}
suggest that this is not the case.   These authors studied Kasner
solutions of M-theory on the moduli spaces with 16 or more
supercharges.   It was found that every solution contained a
singular region which could not be dual transformed into a weakly
coupled string theory, smooth 11 dimensional SUGRA, or even an
F-theory type regime.   In the singular region, the FSB\cite{fsb}
bounds suggest that the Hilbert space describing physics inside a
particle horizon has a small finite dimension, which shrinks as
one approaches the singularity.   It seems unlikely that any
weakly coupled string theory could describe such a situation.

Although these results suggest that singularities cannot be
removed at string tree level, the question remains an interesting
one and deserves further study.

Finally, we would like to suggest an hypothesis about the
relevance of perturbative string calculations to physics in the
FSC background.  String theory is an S-matrix theory, and as such
appears to describe infinite numbers of incoming and outgoing
initial states.  The Fischler-Susskind prescription formally
preserves this property.   On the other hand, we have suggested
that all solutions of the FS equations have at least a Big Bang
singularity.  It has been suggested that there may be a unique
initial state at such a singularity.   One way in which such a
conclusion might arise self consistently within the stringy
formalism is that most scattering amplitudes calculated by the FS
prescription would simply diverge.   One would then discover that
the divergences vanished only for a particular choice of initial
state (and presumably only after a resummation of the perturbation
expansion).

\subsection{\bf Metastable DS and Isolated M Vacua}

There has been much recent interest in string models which
stabilize all moduli at values where string perturbation theory
might be valid.   A basic idea is that nonzero Ramond-Ramond
fluxes on cycles of the compactification manifold, and D-branes
wrapped on such cycles, give contributions to the energy that
scale as different powers of the string coupling.   By
contemplating large fluxes, one can stabilize the dilaton at weak
coupling. It is harder to stabilize the volume of the
compactification manifold , and in fact the best that has been
achieved so far is to generate a no scale model in the SUGRA
approximation\cite{joeetal}\footnote{This statement does not take
into account as yet unpublished material.  J.Polchinski and E.
Silverstein have independently informed us of forthcoming work in
which stable, nonsupersymmetric AdS vacua of weakly coupled string
theory are found.   E. Silverstein has also proposed a more
speculative scenario for describing a metastable DS minimum. The
remarks which follow were written before we learned of
these new results.} . Higher order corrections will give a
potential for the volume modulus, which vanishes at infinite
volume. Perhaps the large fluxes will appear in this potential in
a way that gives it a minimum at a value where systematic
calculations are possible.

Similarly, there are perturbative string compactifications on
asymmetric orbifolds, which freeze all the geometrical moduli,
leaving only the dilaton.  One can imagine {\it e.g.} racetrack
scenarios in which a calculable minimum is found at weak coupling.
Again, the potential will vanish at asymptotically weak coupling.

In all such models we have two candidate background geometries for
string theory.   The first is the FSC solution we have been
discussing in this paper, in which the modulus starts infinitely
far away, rolls up the hill of the potential and rolls back
again.   Although this solution contains a Big Bang singularity,
we have argued that it should be described by a well defined
S-vector.  Although this does not allow us to freely specify
initial conditions, one can certainly imagine that, because the
barriers between the nontrivial minimum and the state at infinite
modulus are parametrically smaller than the Planck scale, there
is a finite probability to push the system into the nontrivial
minimum in some local region in space.  The question now is how
such an event manifests itself in the scattering amplitudes.

The value of the potential at its minimum is clearly an important
determinant of what happens. If it is negative, there is an
instanton\cite{cdl} that describes decay of the FSC into a stable
AdS minimum.   Since we are assuming both the vacuum energy and
the barrier between the AdS minimum and infinity are
parametrically smaller than the Planck scale (defined in terms of
the minimum value the Planck scale of the noncompact dimensions
attains in the FSC solution ), this instanton is below the
Coleman-DeLuccia bound and the decay actually occurs\footnote{The
natural scale of variation of the potential for the moduli is the
string scale, parametrically smaller than the Planck scale.  This
is important to the conclusion about the CDL bound.}.
Furthermore, the expansion of the universe in the FSC solution is
subluminal, so vacuum bubbles collide and percolate. There is,
strictly speaking, no FSC state of the system, which is rather
described by an AdS vacuum of string theory.   By the AdS/CFT
correspondence, this suggests the existence of a nonsupersymmetric
conformal field theory with the rather peculiar pattern of
operator dimensions that are necessary to describe a large radius
AdS space.  If the string coupling at the AdS minimum is small,
one imagines this CFT to be a gauge theory with relatively large N
and large 't Hooft coupling.  An interesting ``inverse question"
arises: is it possible to see evidence for a metastable FSC state
in the large N gauge theory?

If the value of the potential at the minimum is positive, we are
close to the situation investigated by Guth and Farhi \cite{gf}:
we are attempting to create a bubble of DeSitter universe in our
FSC background.  If we work in Einstein frame this situation
resembles that of Guth and Farhi so much that their conclusion
follows.  The stress tensor satisfies the dominant energy
condition.   The analysis of black hole formation is essentially
local, and at least in the late stages of the FSC cosmology, when
the universe is expanding slowly, it is unaffected by the general
cosmological expansion.  Indeed, Guth and Farhi intended their
analysis to apply to the real world, which is a Robertson-Walker
cosmology and not an asymptotically flat universe.   The FSC
cosmology differs from that of the real world only by the
equation of state of the dominant matter at late times.

The conformal factor relating the string frame to the Einstein
frame is singular only asymptotically, when the string coupling
goes to zero in the infinite past and future.  The Guth-Farhi
black hole is formed locally, at a time when the string coupling
is finite.  Thus, again, we cannot invoke the transformation to
string frame to attempt to avoid the conclusions of their
analysis.

We now turn to the case of an hypothetical, isolated
asymptotically flat vacuum which is calculable.  Namely, we assume
that by inserting a number of large fluxes, one stabilizes all
moduli including the dilaton, at a value where string perturbation
theory is applicable and the vacuum energy is exactly zero.
Needless to say, there are no known examples of vacua of this
type.   One may be interested in them in two different contexts:
the first is conventional string phenomenology, where such a
vacuum is presumed to be nonsupersymmetric,  and corresponds to
the real world.  The conjectures of \cite{tbfolly} deny this
possibility but postulate the existence of a supersymmetric
isolated vacuum state toward which the theory of the real world
would asymptote if the cosmological constant were taken to zero.
In either case one must address the interesting practical
question of whether perturbative calculations in string theory
can have any relevance for the real world.  This could be the
case if the isolated vacuum occurred at weak string coupling
because of the existence of large topological invariants like
fluxes.

As above, these scenarios exhibit two classical background
spacetimes. The first is the FS cosmology at asymptotically weak
coupling, while the second is the isolated asymptotically flat
vacuum.   We have to ask which of these is stable, and whether one
can detect the unstable one inside the stable one.  The stability
of the isolated vacuum toward decay into the FSC solution seems
clear. If we use the symmetries of the FSC solution it would seem
that the only possible way to compare the two backgrounds is to
match the cosmological time of the FSC solution to the Minkowski
time in some Lorentz frame.   At any finite cosmological time, the
FSC solution has positive energy density, and there is no
instanton that allows the asymptotically flat spacetime to decay
into it. In the case of an exactly SUSY vacuum, stability follows
from SUSY. SUSY violating, asymptotically Minkowski vacua can
sometimes exhibit semiclassical instability\cite{wfh}, but they
decay into "nothing", rather than into a positive energy
cosmology.

The question of stability of the FSC solution is more subtle. For
simplicity of exposition we will restrict attention to four
dimensions, though the generalization to arbitrary dimensions is
easy. If we neglect the cosmological expansion, it is clear that,
given the presumed parametrically small potential for the dilaton,
there are bubbles of isolated vacuum whose growth is energetically
favored. Let $\epsilon$ be the instantaneous energy density
difference between the Minkowski and FSC solutions and $\sigma$
the instantaneous surface tension of a bubble separating them.
Then, neglecting numbers of order one, the critical bubble size is
${\sigma\over\epsilon}$. Assuming both $\sigma$ and $\epsilon$ are
much smaller than the Planck scale, we would normally presume such
bubbles to expand with the speed of light. Since the cosmological
expansion of the FSC background is subluminal we might then expect
percolation of the bubbles and complete decay of the FSC cosmology
into flat spacetime.

This analysis neglects the time dependence of the parameters of
the bubble.  The energy density difference $\epsilon$ is
constantly decreasing with time. Further, if we consider the late
time evolution, when the dilaton is moving toward the weak
coupling regime and away from the isolated minimum, then $\sigma$
is increasing with time.   The critical bubble size is thus
increasing rapidly with time at late times.  It makes no sense to
discuss a critical bubble of size larger than the cosmological
horizon.  At any time the universe can be viewed as made up of
decoupled quantum systems, which describe physics inside disjoint
backward lightcones whose tips lie on that time slice. Thus, we
must have ${\sigma\over\epsilon} < {M_P \over \sqrt{\epsilon}}$.
As above, $M_P$ is defined in terms of the value of the string
coupling at the turnaround point, and does not vary with time.
Bubble nucleation must surely stop once this inequality fails, and
this is inevitable as the universe expands .

Indeed, it is very likely that the bubble nucleation process is
dominated by events which occur near the point of turnaround, when
the dilaton reaches its maximal height on the potential.   Earlier
on, the expansion rate of the universe is much larger than the
bubble nucleation rate and the Coleman-DeLucia analysis \cite{cdl}
that we have been using is inapplicable. The horizon size is very
small. It is reasonable to presume that few if any bubbles are
nucleated during this period.   Near the turnaround point the
instanton action is at its minimum: the energy density is small
compared to the Planck scale and is of the same order of magnitude
as the surface tension , while the classical FSC configuration is
as close as it gets to the isolated minimum.  Within the realm of
validity of the semiclassical analysis, the probability per unit
time per unit volume for bubble nucleation will be very small.

Now consider the expansion of these bubbles.   The cosmological
increase in the tension of their walls and simultaneous decrease
of $\epsilon$ will act to slow the expansion.  It seems likely to
us, though we have as yet no proof, that the asymptotic expansion
rate is likely to be slower than the speed of light.   The bubble
thus becomes visible to a distant observer, and will have (if it
continues to expand) a radius of order $R > {\sigma\over\epsilon}$
and mass of order $\sigma_{eff} R^2$.  Here $\sigma_{eff}$ is an
effective surface tension.  One might guess a formula
$\sigma_{eff} \sim \sigma \sqrt{1 - v^2}$ where $v$ is the
asymptotic speed of the bubble expansion. It is clear that if the
radius gets too large, the bubble will be inside its own
Schwarzchild radius and will collapse into a black hole. Thus,
there are bounds on the bubble size of order
${\sigma\over\epsilon} < R < {M_P^2 \over \sigma_{eff}}$.  It is
clear that $\sigma_{eff} $ must increase as the universe expands.
It is proportional to $\sigma$, which grows, and the velocity
certainly should not increase as the universe expands.  Thus,
eventually the bubble must recollapse, and the FSC solution is
stable.  We are aware that this argument is far from rigorous and
that we have not completely ruled out the possibility of decay of
the FSC solution into isolated vacuum, but we believe it is
implausible.

We will now briefly discuss the question of whether finite energy
processes in either the isolated Minkowski or FSC backgrounds can
create large metastable bubbles of the other solution, which could
be explored by experimentalists living in one of these alternative
universes.   We begin with the asymptotically Minkowski
background, which has a well defined S-matrix. A bubble of FSC
solution will have finite energy, parametrically smaller than the
Planck scale.  In the setup we are imagining, the barriers between
the FSC regime and the Minkowski vacuum are also parametrically
small.   Thus, there will be a range of bubble sizes for which the
bubble is larger than its Schwarzchild radius.   As time goes on
inside the bubble, the dilaton decreases.  This would tend to
increase the barrier between the FSC and Minkowski solutions, and
therefore must contribute to accelerated collapse of the bubble.
However, given our assumptions, we can tune the rate of these
processes to be slow by tuning large fluxes.   Finally note that
since the expansion inside the bubble is subliminal, there is no
paradox in assuming that an observer dropped into the bubble can
report back to his colleagues outside about the processes going on
there.

Similar remarks apply to creation of a small bubble of Minkowski
vacuum in the FSC cosmology.  There is a small philosophical
difference.   If we truly believe in the notion of an S-vector
rather than an S-matrix, we have to accept the (surely
approximate) notion of the free will of local observers in order
to claim with $100 \%$ probability that such experiments can
actually be done.   If the time evolution of the universe unfolded
uniquely from a unique initial state then one would only have to
hope that there was a sufficiently large probability that a bubble
creation event occurred.  This is to be contrasted to the
Minkowski situation, where an infinite set of initial conditions
is part of the mathematical setup.   Again, with the parameters as
we are assuming them, the bubble creation events are sufficiently
localized and occur at sufficiently low energies that these
unpalatable philosophical questions are probably not important.

At any event, we are not really interested in the bubble creation
experiments, which are totally impractical even if not ruled out
in principle.   The key issue is whether, in the situation we are
hypothesizing, the mathematical apparatus of perturbative string
theory, which (apart from times near the Big Bang singularity)
describes the physics of the FSC solution, can be used to
calculate properties of the isolated Minkowski vacuum.   The above
considerations suggest that this is indeed the case, at least with
some limited accuracy.

\section{What the Dark Energy Isn't, and What it Might Be}

Much of this paper has been devoted to a demonstration that
accelerated expansion does not occur in controllable situations in
string perturbation theory, and to an exploration of the
perturbative physics that does occur. In this section we briefly
discuss stringy perspectives on the problem of Dark Energy.

A number of proposals have been made for the dark energy. While we
can hardly claim to understand what the dark energy might be in
string theory, many of these proposals seems implausible. There
are a variety of difficulties. Some have to do with the required
scales; some are related to the problem of horizons. Still others
have to do with issues peculiar to possible anthropic
explanations.

Among the proposals are a variety having to do with brane
pictures.   In scenarios with large but finite extra dimensions,
standard effective field theory arguments indicate that one will
inevitably obtain too large a cosmological constant.  In theories
with infinite extra dimensions, the effective theory arguments do not
immediately apply, but the various proposals lead to
singularities, whose interpretation is at best unclear.
We have nothing further to add
on this question here.

We have already noted the arguments of \cite{fetal}\cite{hks} that
quintessence, if it is eternal, leads to horizons which are
problematic in string theory.  We will note further difficulties
with quintessence below, associated with the required scales.

One of the most puzzling aspects of the dark energy problem is the
question of coincidence:  why is the scale of the dark energy
today so close to that of the dark matter?  At the moment, the
most plausible explanations of this fact are anthropic.
Inevitably, any successful anthropic explanation of the
cosmological constant problem will predict a dark energy density
within an order of magnitude or so of the dark matter
density\cite{weinberg,vilenkin}.  At least two classes of
anthropic explanations have been widely discussed recently.  The
first requires the presence in the theory of a large number of
possible four form fluxes.  The different discrete values of these
fluxes then lead to a large number of metastable states with a
``discretum" of energies.  There are a number of difficulties with
this proposal, which are discussed in \cite{bdm}.  In particular,
in these schemes, one needs to suppose that there are a vast
number of non-supersymmetric (metastable) states. In string theory
we have yet to reliably exhibit one.  Perhaps more fundamental is
the problem that in these proposals not only is the cosmological
constant determined anthropically, but all of the other parameters
of the standard model are either anthropic or random variables.
But this seems unlikely. While we might imagine that anthropic
considerations would determine the masses of the light quarks and
leptons, for example, it is less plausible that such
considerations determine the heavy quark masses and mixings. These
parameters hardly appear random.

The
second class of anthropic proposals requires
the presence of an extremely light scalar, with
Compton wavelength of order the size of the present horizon or
larger.  The idea is that the value of this field is essentially a
random variable during inflation.  Different parts of the universe
will have different values of the cosmological constant depending
on the value of the field in that region.  If, for example, the
potential is $m^2 \phi^2$, then this can cancel a negative
cosmological constant, say of order $10^8 GeV^4$, provided that
$\phi$ is large enough.

In string theory, however, it seems implausible that a field so
light can carry so much energy.  How, first, might we imagine
getting such a light field?  There is no evidence that in string
theory, scalar fields are appreciably lighter than the scale of
supersymmetry breaking, in the absence of a symmetry.  More
precisely, there are many situations where we can study
supersymmetry breaking in a controlled approximation.  In these
cases, all fields gain mass of order the supersymmetry breaking
scale, $M_{\small susy}$, or perhaps $M_{\small susy}^2 \over
M_p$.  This is much the same as occurs in supergravity theories
\footnote{Note that in the case of moduli, in known, controlled
examples, the moduli have non-trivial potentials; by mass we mean
the second derivatives of the potential, where appropriate.}.

The only examples of symmetries which might yield such light
scalars are axions.  Now the mass of an axion might plausibly be
of order
\beq m_a^2 = e^{-{8 \pi^2 \over g^2}} M_{susy}^4/M_p^2 \eeq where
we might imagine that $g$ is of order some typical unified
coupling, and supposed that the axion decay constant is within a
few orders of magnitude of the Planck mass. We might also imagine
that $M_{susy} \sim 10^{10} {\rm GeV}.$  This would give \beq m_a
\sim 10^{-33} GeV \eeq which gives a Compton wavelength not
wildly different than the size of the universe (it is about three
orders of magnitude smaller).  Given the huge uncertainties in
this estimate, this is an interesting result.

In any case, the potential for such an axion is periodic, with
period approximately $f_a$, and one
does not expect that it will be larger than
\beq
V_o =  e^{-8 \pi^2 \over g^2} M_{susy}^4
\eeq
i.e. it will be of order 70 orders of magnitude smaller than the
expected contributions to the vacuum energy from supersymmetry
violation.

It is interesting, on the other hand, that this crude estimate is
in the right ballpark for the axion itself to provide the dark
energy.  There are two ways this might happen.  First, we might
postulate that, in addition to the axion which explains the
smallness of the QCD theta parameter, there is another axion, with
mass of order the mass given by this estimate.  Then this axion
might still be frozen at a point away from its minimum, and the
observed dark energy could just be this stored
energy\cite{carroll,wittenaxion}. For this to be the case,
however, it is important that the axion energy density should be
of order the observed energy density, while the mass is small.
This is problematic. It requires an additional coincidence: in
order that the axion not be rolling now (so that it's equation of
state will resemble that of a cosmological constant) it is
necessary that its mass be smaller than the present Hubble
constant.  But this mass is related to the energy density by
(assuming that the axion energy is a simple cosine, i.e. $V= C
cos(a f_a)$, or similar periodic function) \beq m_a^2 f_a^2 =  ({3
\over 8 \pi}) H^2 M_p^2 \eeq where $a_o$ is the present value of
the axion field. So even if the axion decay constant is as large
as the (reduced) Planck mass, the axion compton wavelength will
not be larger than the present horizon. So in effect, we now have
two coincidences: the potential is just such that it dominates the
energy density during the current epoch, and the axion decay
constant is just such that the axion is about to roll, but hasn't
quite begun yet. \footnote{Note, for example, that in the
Horava-Witten picture, the axion decay constant is several orders
of magnitude below the Planck mass; in weakly coupled string
theory, it is of order the string scale, which is suppressed by a
factor of coupling relative to the Planck scale. It seems unlikely
that one can obtain a decay constant much larger than the Planck
scale}.

So it seems unlikely that the explanation of the dark energy is
that there is an axion sitting near the top of a hill.  This
requires a particle present solely for this purpose, with both
energy density and decay constant (mass) tuned just so.

Similar remarks apply to quintessence, which also requires a field
with a Compton wavelength comparable to the present horizon, but
whose energy density must be comparable to the present energy
density.   Once more, in the absence of a symmetry, we know of no
example in string theory where the scale of the potential of a
particle is not related to the scale of supersymmetry breaking,
without some additional, Peccei-Quinn like symmetry. Difficulties
with axion-like particles as quintessence, beyond those described
above, have been discussed in \cite{choi,carroll}.  The former
reference outlines in some detail the special circumstances
required to obtain axion domination now.  It points out
difficulties with fields other than the axion as quintessence.
The latter discusses observational difficulties associated with
a quintessence axion.

There is a possible alternative role for axions, which doesn't
require the addition of a particle solely for the purpose of
explaining the dark energy, and which requires only one, not
totally implausible, coincidence.  Consider the ordinary QCD
axion.  As has frequently been discussed in the axion literature,
if the coefficient of the QCD anomaly is not $1 \over 16\pi^2$ but $N
\over 16 \pi^2$, then the QCD contribution to the
axion potential takes the form \beq V=
m_{\pi}^2 f_{\pi}^2 cos(a N f_a) \eeq In this case, there are $N$
degenerate ground states.  This degeneracy holds exactly in QCD,
and reflects the fact that $QCD$ breaks the original PQ symmetry
down to a $Z_N$.  More precisely, it holds in the limit that
\begin{itemize}
\item
Only effects connected with the anomaly break the
Peccei-Quinn symmetry.
\item
The SU(2) gauge coupling is set to zero.
\end{itemize}

Once we turn on the $SU(2)$ gauge coupling, the $Z_N$ symmetry
may be broken by $SU(2)$ instantons.  This will be the case if the
anomalous coupling of the axion to $SU(2)$ is different than to
$SU(3)$ (say $1$ instead of $N$).  Considerations of the low
energy, renormalizable theory might suggest that these
contributions will receive additional suppression, involving many
Yukawa couplings and loop factors.  These are necessary to tie
together the many fermion zero modes.  But in general it should
be possible to tie up these zero modes with high dimension
operators.  Indeed, no symmetry (except, possibly, anomalous
discrete symmetries) can forbid the appearance of high dimension
operators with the quantum numbers of the 't Hooft operator itself
under any approximate low energy symmetries.  These operators
will be suppressed by powers of the large scale (e.g. the
unification scale). This just means that the principle
contribution to the amplitude will come from very small
instantons.   The real suppression lies in the exponential of the
gauge coupling, but by using the unified coupling in our
estimates, we have taken this into account. Additional
suppression factors, such as powers of $\pi$ etc., depend on the
details of the theory.

More generally, then, we might expect, due to instantons at the
unification scale, $M$, symmetry breaking effects of order \beq
e^{-8 \pi^2\over g^2(M)} M_{susy}^4 \eeq and, as we have seen,
this is a number easily within a few orders of magnitude of the
observed dark energy density!  In other words, we might imagine
that in the lowest energy state, for (mysterious) reasons, the
cosmological constant vanishes; then there are a set of nearly
degenerate states, with an energy density of order that which is
observed!\footnote{As this paper was being completed, we received
the paper \cite{barr} which also argues that the lifting of the
$N$-fold axion degeneracy might account for the dark energy.  In
this proposal, it is argued that operators of very high dimension
might break the PQ symmetry by a tiny amount, small enough not to
spoil the solution of the strong CP problem, but large enough to
account for the vacuum energy.}

Of course, in this view, the cosmic coincidence seems to be an
accident, with a chance of order one part in a thousand, or
perhaps smaller.  We would note, however, that this is no worse
than another, somewhat more vague suggestion for understanding the
coincidence\cite{halletal}.  Some authors have noted that that the
observed dark energy density is very nearly the fourth power of
${\rm TeV}^2 \over \tilde M_p$, where $\tilde M_p$ is the reduced
Planck mass.  They have argued that this is a plausible form for a
microscopic expression for the energy density, given that $TeV$ is
of order the weak scale. Indeed, this gives a result within a
factor of 10 of the observed density.  On the other hand, writing
the formula as \beq \Lambda = {(c~{\rm TeV})^8 \over M_p^4} \eeq
makes clear that there are many orders of magnitude uncertainty
even in this crude estimate.  For example, if $c$ is $3$; this
would increase the answer by a factor of almost $10^4$!  So we
would claim our proposal is as good (or bad) an explanation of the
cosmic coincidence as any other non-anthropic proposal.

Finally we mention the proposal of \cite{tbfolly}, to which one of us
must confess a certain attachment.   In this proposal, there is a true
cosmological constant.   Furthermore, it is assumed to be a fundamental
input parameter, rather than a calculable quantity in the low energy
effective action.   The reasoning is that the cosmological constant,
according to the holographic principle, measures the total number of
states in the Hilbert space describing the universe.   In quantum
mechanics, the total number of states is always a fixed boundary
condition, rather than a dynamical quantity.

>From this point of view, the puzzle of the actual value of
$\Lambda$ would be resolved only by anthropic reasoning.   There
could be a Meta-theory that produces some probability
distribution for the size of the Hilbert space describing a
particular universe\footnote{We are quite uneasy about the
prospect of such a theory.  It describes the probabilities for
alternative universes and in principle can never be tested in our
own.  Furthermore, as long as the probability distribution it
produces has nonzero support in the anthropic region and is not
peaked at one of the extremes of that region it will be
compatible with observations. Thus, few if any of the details of
this theory have any effect on any observable quantity.}

The other possibility is that the number of states $N$ has to
satisfy some number theoretic identity whose solutions are very
sparse.   At first sight it would appear that the number
$e^{10^{123}}$ is so huge that it is hard to believe this
possibility.   On the other hand, there are problems in number
theory which have no or only a few known solutions, but no proof
to date that there is no other\footnote{The simple example is:
find an odd perfect number.}.  Perhaps, for some peculiar reason,
the number of states has to be an odd perfect number. Another
possibility, which does not rely on an unproven mathematical
conjecture is that the number of states has to be of the form
$2^p$ where $p$ is a Mersenne prime (a prime of the form $2^k -
1$). There are only two values of $k$, $521$ and $607$, in the
table of Mersenne primes which would give a cosmological constant
within shouting distance of the true value (and both are off by
many orders of magnitude) according to this formula. Thus,
although the particular example of Mersenne primes does not work,
it is easy to imagine number theoretic criteria that would allow
only one value of the number of states which was in any way
realistic. Systems with vastly smaller numbers of states could not
exhibit any sort of interesting physics, while those with vastly
larger numbers of states are likely to be described by low energy
physics that is superconformally invariant down to extremely low
energy scales \cite{tbfolly}.

In the end, the number theoretic option still has to resort to the
anthropic principle.
However, (assuming that it is easy to rule out life in a superconformal
world) the anthropic arguments are much simpler.   For all solutions
except one the world would either be described by a system with a number
of states too small to support complex systems, or a system that was
superconformal down to incredibly low energy scales.

Of course the real challenge for this set of ideas is
verification of the claim that SUSY breaking scales with an
unconventional power of $\Lambda$ as $\Lambda \rightarrow 0$.

\noindent {\bf Acknowledgements:}

\noindent
We thank R. Bousso, W. Fischler, N. Kaloper,
J.Polchinski, E.Silverstein and L.Susskind for discussions. This
work was supported in part by the U.S. Department of Energy.


\newpage
\begin{thebibliography}{19}        


\bibitem{tbf} T.Banks, W.Fischler, {\it M-theory Observables for
Cosmological Space Times}, hep-th/0102077.
\bibitem{fetal} W.Fischler, A.Kashani-Poor,  R.McNees, S.Paban,
{\it The Acceleration of the Universe, A Challenge For String
Theory}, hep-th/0104181.
\bibitem{hks} S.Hellerman, N.Kaloper, L.Susskind, {\it String
Theory and Quintessence}, JHEP 0106, 003, 2001, hep-th/0104180.
\bibitem{acc}
For recent developments, see for example:
A. Reiss et al, {\it The Farthest Known Supernova:  Support for an
Accelerating Universe and a Glimpse of teh Epoch of Deceleration,}
astro-ph/0104455.
\bibitem{fs} W.Fischler, L.Susskind, {\it Dilaton Tadpoles, String
Condensates and Scale Invariance, 1 and 2}, Phys. Lett. B171,
383, (1986); Phys. Lett. B173, 262, (1986).
\bibitem{bs} R.Brustein, P.Steinhardt, {\it Challenges For
Superstring Cosmology}, Phys. Lett. B302, 196, (1993),
hep-th/9212049.
\bibitem{w} E.Witten, Talk at Strings 2001, Mumbai, India.
\bibitem{joeetal} S.Giddings, S.Kachru, J.Polchinski, {\it
Hierarchies From Fluxes in String Compactifications},
hep-th/0105097.
\bibitem{gf} E.Farhi, A.Guth, {\it An Obstacle to Creating a
Universe in the Laboratory}, Phys. Lett. B183, 149, (1987).
\bibitem{isovac} T.Banks, {\it On Isolated Vacua and Background
Independence}, hep-th/0011255.
\bibitem{thsut} C.R. Stephens, G.'tHooft, B.Whiting,
{\it Black Hole Evaporation Without Information Loss},
Class.Quant.Grav. 11, 621, (1994), gr-qc9310006; L.Susskind,
L.Thorlacius, J.Uglum, {\it The stretched horizon and black hole
complementarity}, Phys. Rev. D48, (1993), 3743, hep-th/9306069.
\bibitem{tbfolly} T.Banks, {\it Cosmological Breaking of
Supersymmetry}, hep-th/0007146
\bibitem{gmh} G.Moore, J.Horne, {\it Chaotic Coupling Constants},
Nucl. Phys. B432, 109, (1994), hep-th/9403058.
\bibitem{bbmss} T.Banks, M.Berkooz, G.Moore, S.Shenker,
P.Steinhardt, {\it Modular Cosmology}, Phys. Rev. D52, 3548,
(1995), hep-th/9503114.
\bibitem{singularities}
The singularity theorems are discussed, for example, in
R. Wald, {\it General Relativity}, University of Chicago Press
(Chicago) 1984.
\bibitem{bfm} T.Banks, W.Fischler, L.Motl, {\it Dualities vs.
Singularities}, JHEP 9901, 019, 1999, hep-th/9811194; T.Banks,
L.Motl, {\it On The Hyperbolic Structure of Moduli Spaces of
M-theory With Sixteen Supercharges}, JHEP 9905,015,1999,
hep-th/9904008.
\bibitem{fsb} W.Fischler, L.Susskind, {\it Holography and Cosmology}, hep-th/9806039.;
R.~Bousso{\it A Covariant Entropy Conjecture },
JHEP 9907 (1999) 004, hep-th/9905177 ;{\it Holography in General
Space Times}, JHEP 9906 (1999) 028, hep-th/9906022; {\it The
Holographic Principle for General Backgrounds},Class.Quant.Grav.
17 (2000) 997, hep-th/9911002
\bibitem{cdl} S.Coleman, F. DeLuccia, {\it Gravitational Effects
On and Of Vacuum Decay}, Phys. Rev. D21, 3305, (1980).
\bibitem{wfh} E.Witten, {\it Instability of the Kaluza-Klein Vacuum},
Phys. Lett. B289, 293, (1992), hep-th/9203031; M.Fabinger,
P.Horava, {\it Casimir Effect Between World Branes in Heterotic M
Theory}, Nucl. Phys. B580, 243, (2000), hep-th/0002073.
\bibitem{weinberg} S. Weinberg, Phys. Rev. Lett. {\bf 59} (1987) 2607;
{\it The Cosmological Constant
Problems}, astro-ph/0005265.
\bibitem{vilenkin} A. Vilenkin, {\it Predictions from Quantum Cosmology},
Phys. Rev. Lett. {\bf 74} (1995)
846, gr-qc/9406010.
\bibitem{bdm} T.Banks, M.Dine, L.Motl, {\it On Anthropic Solutions
to the Cosmological Constant Problem}, JHEP 0101,031,2001,
hep-th/0007206.
\bibitem{wittenaxion}
E. Witten, {\it The Cosmological Constant from the Viewpoint of
String Theory,} hep-ph/0002297.
\bibitem{barr}
S. Barr and D. Seckel, {\it The Cosmological Constant, False Vaua and
Axions,}, hep-ph/0106239.

\bibitem{halletal} N.Arkani-Hamed, L.J.Hall, C.Kolda, H.Murayama,
Phys. Rev. Lett. 85, 4434, (2000), astro-ph/0005111.

\bibitem{choi}
K. Choi, {\it String or M Theory Axion as a Quintessence,}
Phys.Rev. {\bf D62} (2000) 043509,
hep-ph/9902292.

\bibitem{carroll} S.Carroll, {\it Quintessence and The Rest of The
World}, Phys. Rev. Lett. 81, 3067, (1998), astro-ph/9806099.
\end{thebibliography}
\end{document}